\documentclass [11pt,a4paper] {article}

\usepackage[cp1252]{inputenc}

\usepackage{amssymb}
\usepackage{amsmath}
\usepackage{amsfonts,amssymb}
\usepackage[dvips]{graphicx}
\usepackage{bbm}
\usepackage{enumerate}

\DeclareMathAlphabet{\mathpzc}{OT1}{pzc}{m}{it}

\def\SmallColSep{\setlength{\arraycolsep}{1pt}}

\begin{document}

\title{Quantum contextuality implies a logic that does not obey the principle of bivalence}

\author{Arkady Bolotin\footnote{$Email: arkadyv@bgu.ac.il$\vspace{5pt}} \\ \textit{Ben-Gurion University of the Negev, Beersheba (Israel)}}

\maketitle

\begin{abstract}\noindent In the paper, a value assignment for projection operators relating to a quantum system is equated with assignment of truth-values to the propositions associated with these operators. In consequence, the Kochen-Specker theorem (its localized variant, to be exact) can be treated as the statement that a logic of those projection operators does not obey the principle of bivalence. This implies that such a logic has a gappy (partial) semantics or many-valued semantics.\\

\end{abstract}

\section{Introduction}  

\noindent Consider the triple $(\mathcal{H},|\Psi\rangle,\mathcal{O})$ in which $\mathcal{H}$ is the Hilbert space of a quantum system, $|\Psi\rangle \in \mathcal{H}$ is the normalized vector describing the state of this system, and $\mathcal{O} = \{\hat{P}\}$ is a finite set of projection operators $\hat{P}$ on $\mathcal{H}$.\\

\noindent Define \textit{an assignment function} $h$ as a function from the set $\mathcal{O}$ to the set of numerical values $\{0,1\}$, namely,\smallskip

\begin{equation}  
      h:
      \mathcal{O}
      \rightarrow
      \{0,1\}
      \;\;\;\;  ,
\end{equation}
\smallskip

\noindent such that\smallskip

\begin{equation}  
      h(\hat{0})
      =
      0
      \;\;\;\;  ,
\end{equation}

\begin{equation}  
      h(\hat{1})
      =
      1
      \;\;\;\;  ,
\end{equation}
\smallskip

\noindent where $\hat{0}$ and $\hat{1}$ are the zero-projection and identity-projection operators, respectively. According to \cite{Abbott12}, the assignment function $h$ expresses the notion of \textit{a hidden variable}, namely, $h(\hat{P})$ specifies in advance the result obtained from the measurement of an observable corresponding to the particular projection operator $\hat{P}$.\\

\noindent Also, define a subset $\mathcal{C} \subset \mathcal{O}$ as \textit{a context} if $|\mathcal{C}| \ge 2$ and any two projection operators from $\mathcal{C}$, say, $\hat{P_i}$ and $\hat{P_j}$, are orthogonal\smallskip

\begin{equation}  
      \hat{P_i}, \hat{P_j} \in  \mathcal{C},
      i \neq j
      \,
      \implies
      \,
      \hat{P_i}
      \hat{P_j}
      =
      \hat{P_j}
      \hat{P_i}
      =
      \hat{0}
      \;\;\;\;  .
\end{equation}
\smallskip

\noindent The context $\mathcal{C}$ is \textit{maximal} (or \textit{complete}) if the projection operators from $\mathcal{C}$ resolve to the identity-projection operator:\smallskip

\begin{equation}  
      \sum_{\hat{P_i} \,\in\,\mathcal{C}}
      \hat{P_i}
      =
      \hat{1}
      \;\;\;\;  .
\end{equation}
\smallskip

\noindent Now consider the following sentences:\smallskip

\begin{enumerate}[(a)]
\item The set $\mathcal{O}$ is value definite under $h$; in other words, $h$ is a total function.
\item 	The value $h(\hat{P_i})$ depends only on $\hat{P_i}$ and not the context $\mathcal{C}$ containing $\hat{P_i}$.
\item 	For a maximal $\mathcal{C}$, the values of its projection operators $h(\hat{P_i})$ add up to 1; otherwise stated, the next entailment holds:

\begin{equation}  
      h
      \left(
         \sum_{\hat{P_i} \,\in\,\mathcal{C}}
         \hat{P_i}
         \!
      \right)
      \!\!
      =
      \!
      1
      \;
      \implies
      \;
      \sum_{\hat{P_i} \,\in\,\mathcal{C}}
      h
      \left(
         \hat{P_i}
      \right)
      \!
      =
      \!
      1
      \;\;\;\;  .
\end{equation}
\end{enumerate}
\bigskip

\noindent Provided that the sentences (b) and (c) are true, the sentence (a) must be denied in accordance with the Kochen-Specker theorem \cite{Kochen, Peres}. That is, the assignment function $h$ \textit{cannot be total} and, hence, at least one projection operator from $\mathcal{O}$ must be value indefinite under $h$ (i.e., must have the value neither 0 nor 1). What is more, according to \textit{the variant of the Kochen-Specker theorem localizing value indefiniteness} \cite{Abbott15}, there is a set $\mathcal{O}$ containing projection operators $\hat{P_i}$ and $\hat{P_j}$ such that if the system is prepared in the pure state $|\Psi_i\rangle$ in which $h(\hat{P_i})=1$, then both $h(\hat{P_j})=1$ and $h(\hat{P_j})=0$ lead to contradictions.\\

\noindent On the other hand, consider \textit{a truth-value assignment function} $v_C$ that denotes \textit{a truth valuation in a circumstance} $C$, that is, a mapping from some subset of propositions $\mathcal{P} \subseteq\{\diamond\}$ related to the quantum system (where the symbol $\diamond$ stands for any proposition, compound or simple) to the set of truth-values $\{0,1\}$ (where the value 0 represents ``false'' and the value 1 represents ``true'') relative to a circumstance of valuation indicated by $C$ (such a circumstance can be, for example, the state $|\Psi\rangle$ in which the system is prepared or found):\smallskip

\begin{equation}  
      v_C:
      \mathcal{P}
      \rightarrow
      \{0,1\}
      \;\;\;\;  .
\end{equation}
\smallskip

\noindent Commonly, it is written using the double-bracket notation, namely, $v_C(\diamond) = {[\![ \diamond ]\!]}_C$. The truth-value assignment function $v_C$ expresses the notion of \textit{not-yet-verified truth values}: It specifies in advance the truth-value obtained from the verification of the proposition $\diamond$.\\

\noindent Let the following valuational axiom hold true\smallskip

\begin{equation}  
      v_C
      (
         \hat{P}_{\diamond}
      )
      =
      {[\![ \diamond ]\!]}_C
      \;\;\;\;  ,
\end{equation}
\smallskip

\noindent where $\hat{P}_{\diamond}$ is the projection operator uniquely (i.e., one-to-one) associated with the proposition $\diamond \in \{\diamond\}$.\\

\noindent Assume that the function $h$ coincides with the function $v_C$. Then, the localized variant of the Kochen-Specker theorem is equivalent to the statement that a logic defined as the relations between projection operators $\hat{P}_{\diamond}$ on $\mathcal{H}$ does not obey \textit{the principle of bivalence} (according to which a proposition must be either true or false \cite{Beziau}). In other words, a logic of the projection operators $\hat{P}_{\diamond}$ has a non-bivalent semantics, e.g., a gappy one (in which the function $v_C$ is partial and thus some propositions may have absolutely no truth-value) or a many-valued one (in which there are more than two truth-values).\\

\noindent Let us demonstrate this equivalence in the presented paper.\\

\section{Truth-value assignment for projection operators}  

\noindent Consider the lattice $\mathcal{L}(\mathcal{C})$ formed by \textit{the column spaces} (a.k.a. \textit{ranges}) of the projection operators $\hat{P_i} \in \mathcal{C}$, the closed subspaces of the Hilbert space $\mathcal{H}$. Let the lattice operation \textit{meet} $\wedge$ correspond to the intersection of the column spaces, while the lattice operation \textit{join} $\vee$ correspond to the smallest closed subspace of $\mathcal{H}$ containing their union. Let the lattice $\mathcal{L}(\mathcal{C})$ be bounded, i.e., let it have the greatest element $\mathrm{ran}(\hat{1})=\mathcal{H}$ and the least element $\mathrm{ran}(\hat{0})=\{0\}$.\\

\noindent One can define the lattice operations on $\mathcal{L}(\mathcal{C})$ as follows:\smallskip

\begin{equation} \label{one} 
   \mathrm{ran}(\hat{A})
   \wedge
   \mathrm{ran}(\hat{B})
   =
   \mathrm{ran}(\hat{A})
   \cap
   \mathrm{ran}(\hat{B})
   =
   \mathrm{ran}(\hat{A}\hat{B})
   \;\;\;\;  ,
\end{equation}

\begin{equation}  
   \mathrm{ran}(\hat{A})
   \vee
   \mathrm{ran}(\hat{B})
   =
   \left(
      \left(
         \mathrm{ran}(\hat{A})
      \right)^{\bot}
      \cap
      \left(
         \mathrm{ran}(\hat{B})
      \right)^{\bot}
   \right)^{\bot}
   \;\;\;\;  ,
\end{equation}
\smallskip

\noindent where $\mathrm{ran}(\hat{A}), \mathrm{ran}(\hat{B}) \in \mathcal{L}(\mathcal{C})$ and $(\cdot)^{\bot}$ stands for the orthogonal complement of $(\cdot)$. Given that the orthogonal complement of the column space is \textit{the null space} (a.k.a. \textit{kernel}), that is,\smallskip

\begin{equation}  
   \left(
      \mathrm{ran}(\hat{A})
   \right)^{\bot}
   =
   \mathrm{ker}(\hat{A})
   =
   \mathrm{ran}(\neg\hat{A})
   \;\;\;\;  ,
\end{equation}
\smallskip

\noindent where\smallskip

\begin{equation}  
      \neg\hat{A}
      =
      \hat{1} - \hat{A}
      \;\;\;\;   
\end{equation}
\smallskip

\noindent is understood as negation of $\hat{A}$ such that\smallskip

\begin{equation}  
   \mathrm{ran}(\hat{A})
   +
   \mathrm{ran}(\neg\hat{A})
   =
   \mathrm{ran}(\hat{1})
   =
   \mathrm{ran}(\hat{A} + \neg\hat{A})
   \;\;\;\;  ,
\end{equation}
\smallskip

\noindent it holds that\smallskip

\begin{equation}  
   \mathrm{ran}(\hat{A})
   \vee
   \mathrm{ran}(\hat{B})
   =
   \left(
      \mathrm{ran}(\neg\hat{A}\neg\hat{B})
   \right)^{\bot}
   =
   \mathrm{ran}(\hat{A} + \hat{B})
   \;\;\;\;  .
\end{equation}
\smallskip

\noindent As the closed subspaces $\mathrm{ran}(\hat{P_i})$ and $\mathrm{ran}(\hat{P_j})$ where $\hat{P_i} \perp \hat{P_j}$, i.e., $\hat{P_i}\hat{P_j}=\hat{0}$, are orthogonal to each other, one finds\smallskip

\begin{equation} \label{two} 
   \mathrm{ran}(\hat{P_i})
   \wedge
   \mathrm{ran}(\hat{P_j})
   =
   \mathrm{ran}(\hat{P_i})
   \cap
   \mathrm{ran}(\hat{P_j})
   =
   \{0\}
   \;\;\;\;  .
\end{equation}
\smallskip

\noindent Next, let us consider the truth-value assignments of the projection operators from the lattice $\mathcal{L}(\mathcal{C})$.\\

\noindent Given that $\mathrm{ran}(\hat{1}) = \mathcal{H}$, any arbitrary state of the system $|\Psi\rangle \in \mathcal{H}$ resides in the column space of the identity-projection operator, i.e., $|\Psi\rangle \in \mathrm{ran}(\hat{1})$. But then, being in $\mathrm{ran}(\hat{P_\diamond})$ means $\hat{P_\diamond}|\Psi\rangle =|\Psi\rangle$; so, in agreement with the eigenstate assumption \cite{Abbott12}, one can presume that the function $v_{|\Psi\rangle}$ assigns the truth value 1 to the projection operator $\hat{1}$ in any admissible state of the system $|\Psi\rangle \in \mathcal{H}$.\\

\noindent At the same time, any admissible state of the system $|\Psi\rangle \in \mathcal{H}$ also resides in the null space of the null-projection operator, i.e., $|\Psi\rangle \in \mathrm{ran}(\neg\hat{0})$. It gives $\hat{0}|\Psi\rangle =0\cdot|\Psi\rangle$, consequently, one can presume that the function $v_{|\Psi\rangle}$ assigns the truth value 0 to $\hat{0}$ in any admissible state of the system $|\Psi\rangle \in \mathcal{H}$. This can be written down as\smallskip

\begin{equation}  
   |\Psi\rangle
   \in
   \left\{
      \begin{array}{r}
         \mathrm{ran}(\hat{1}) = \mathcal{H}\\
         \mathrm{ran}(\neg\hat{0}) = \mathcal{H}
      \end{array}
   \right.
   \iff
   \left\{
      \begin{array}{l}
         v_{|\Psi\rangle}(\hat{1}) = 1\\
         v_{|\Psi\rangle}(\hat{0}) = 0
      \end{array}
   \right.
   \;\;\;\;  .
\end{equation}
\smallskip

\noindent Let the system be prepared in a pure state $|\Psi_i\rangle$ lying in the column space of the projection operator $\hat{P_i} \in \mathcal{C}$. Since $\hat{P_i}|\Psi_i\rangle =|\Psi_i\rangle$, one can assume that the function $v_{|\Psi_i\rangle}$ assigns the truth value 1 to the projection operator $\hat{P_i}$ in the state $|\Psi_i\rangle$. Contrariwise, if the truth value of the projection operator $\hat{P_i}$ is 1 in the state $|\Psi_i\rangle$, one can deduce that the state $|\Psi_i\rangle$ is in the column space of the projection operator $\hat{P_i}$. These two suppositions can be recorded together as the following logical biconditional:\smallskip

\begin{equation} \label{three} 
   |\Psi_i\rangle
   \in
   \mathrm{ran}(\hat{P_i})
   \;\;
   \iff
   \;\;
   v_{|\Psi_i\rangle}(\hat{P_i})
   =
   1
   \;\;\;\;  .
\end{equation}
\smallskip

\noindent In view of (\ref{two}), the the vector $|\Psi_i\rangle$ must also reside in the null space of any other projection operator $\hat{P_j}$ in the context $\mathcal{C}$, and therefore all other truth values $v_{|\Psi_i\rangle}(\hat{P_j})$ relating to the context $\mathcal{C}$ must be zero:\smallskip

\begin{equation} \label{four} 
   |\Psi_i\rangle
   \in
   \mathrm{ran}(\neg\hat{P_j})
   \;\;
   \iff
   \;\;
   v_{|\Psi_i\rangle}(\hat{P_j})
   =
   0
   \;\;\;\;  .
\end{equation}
\smallskip

\noindent This obviously gives\smallskip

\begin{equation}  
   \underbrace{
      v_{|\Psi_i\rangle}(\hat{P_i})
      }_{\text{=1}}
   +
   \underbrace{
   \sum_{j \neq i}
   v_{|\Psi_i\rangle}(\hat{P_j})
      }_{\text{=0}}
   =
   1
   \;\;\;\;  .
\end{equation}
\smallskip

\noindent By contrast, consider the state $|\Phi_i\rangle$ where $v_{|\Phi_i\rangle}(\hat{P_i}) = 0$ implying $|\Phi_i\rangle \in \mathrm{ran}(\neg\hat{P_i})$. According to (\ref{one}), in the maximal context $\mathcal{C}$ there is another $\hat{P_k}$, $k \neq i$, such that\smallskip

\begin{equation}  
   \mathrm{ran}(\neg\hat{P_i})
   \cap
   \mathrm{ran}(\hat{P_k})
   =
   \mathrm{ran}(\neg\hat{P_i}\hat{P_k})
   =
   \mathrm{ran}\left((\hat{1} - \hat{P_i})\hat{P_k}\right)
   =
   \mathrm{ran}(\hat{P_k})
   \;\;\;\;  ,
\end{equation}
\smallskip

\noindent and so\smallskip

\begin{equation}  
   |\Phi_i\rangle
   \in
   \mathrm{ran}(\neg\hat{P_i})
   \subseteq
   \mathrm{ran}(\hat{P_k})
   \;\;
   \iff
   \;\;
   v_{|\Psi_i\rangle}(\neg\hat{P_i})
   =
   v_{|\Psi_i\rangle}(\hat{P_k})
   =
   1
   \;\;\;\;  ,
\end{equation}

\begin{equation}  
   \underbrace{
      v_{|\Phi_i\rangle}(\hat{P_i})
      }_{\text{=0}}
   +
   \underbrace{
      v_{|\Phi_i\rangle}(\hat{P}_k)
      }_{\text{=1}}
   +
   \underbrace{
      \sum_{\j \neq i,\,k}               
      v_{|\Phi_i\rangle}(\hat{P_j})
      }_{\text{=0}}
   =
   1
   \;\;\;\;  .
\end{equation}
\smallskip

\noindent Subsequently, if the system is prepared (found) in the state lying in the column or null space of any projection operator from the maximal context $\mathcal{C}$, then among all the propositions $\mathcal{P}_{\mathcal{C}} = \{\diamond\}_\mathcal{C}$ associated with $\mathcal{C}$ exactly one would be true while the others would be false.\\

\noindent Assume that there is a different context $\mathcal{C}^\prime \subset \mathcal{O}$, where some members $\hat{P_i^\prime} \in \mathcal{C}^\prime$ do not commute with $\hat{P_i} \in \mathcal{C}$. Suppose that the state $|\Omega\rangle$ is arranged in the subspace from the lattice $\mathcal{L}(\mathcal{C}^\prime)$, e.g., $|\Omega\rangle \in \mathrm{ran}(\hat{P_l^\prime})$, entailing $v_{|\Omega\rangle}(\hat{P_l^\prime}) = 1$ and $v_{|\Omega\rangle}(\neg\hat{P_l^\prime}) = 0$.\\

\noindent Let us show that the vector $|\Omega\rangle$ resides in neither the column space nor the null space of at least one projection operator, say, $\hat{P_l}$, from the other lattice $\mathcal{L}(\mathcal{C})$ and, as a result, the truth-value function $v_{|\Omega\rangle}(\hat{P_l})$ must assign neither 1 nor 0 to this operator under the valuations (\ref{three}) and (\ref{four}), that is,\smallskip

\begin{equation}  
   |\Omega\rangle
   \notin
   \left\{
      \begin{array}{r}
         \mathrm{ran}(\hat{P_l})
      \\
         \mathrm{ran}(\neg\hat{P_l})
      \end{array}
   \right.
   \iff
   v_{|\Omega\rangle}(\hat{P_l})
   \notin
   \{0,1\}
   \;\;\;\;  .
\end{equation}
\smallskip

\section{Cabello's set of 4{$\times$}4 matrices}  

\noindent Consider the projection operators $\hat{P}^{(1)}_i \in \mathcal{C}^{(1)}$ and $\hat{P}^{(6)}_i \in \mathcal{C}^{(6)}$ on the Hilbert space $\mathcal{H} = \mathbb{C}^{4}$ from the set $\mathcal{O} = \{\mathcal{C}^{(Q)}\}_{Q=1}^9$ of 18 four-dimensional matrices used in the paper \cite{Cabello} by Cabello et al. to prove the Bell-Kochen-Specker theorem:\smallskip

\begin{equation}  
   \hat{P}^{(1)}_1
   =
   \!\left[
      \begingroup\SmallColSep
      \begin{array}{r r r r}
         0 & 0 & 0 & 0 \\
         0 & 0 & 0 & 0 \\
         0 & 0 & 0 & 0 \\
         0 & 0 & 0 & 1
      \end{array}
      \endgroup
   \right]
   \,
   ,
   \,\,
   \hat{P}^{(1)}_2
   =
   \!\left[
      \begingroup\SmallColSep
      \begin{array}{r r r r}
         0 & 0 & 0 & 0 \\
         0 & 1 & 0 & 0 \\
         0 & 0 & 0 & 0 \\
         0 & 0 & 0 & 0
      \end{array}
      \endgroup
   \right]
   \,
   ,
   \,\,
   \hat{P}^{(1)}_3
   =
   \!
   \frac{1}{2}
   \!\left[
      \begingroup\SmallColSep
      \begin{array}{r r r r}
         1 & 0 & 1 & 0 \\
         0 & 0 & 0 & 0 \\
         1 & 0 & 1 & 0 \\
         0 & 0 & 0 & 0
      \end{array}
      \endgroup
   \right]
   \,
   ,
   \,\,
   \hat{P}^{(1)}_4
   =
   \!
   \frac{1}{2}
   \!\left[
      \begingroup\SmallColSep
      \begin{array}{r r r r}
                  1 & 0 & \bar{1} & 0 \\
                  0 & 0 & 0          & 0 \\
         \bar{1} & 0 & 1          & 0 \\
                  0 & 0 & 0          & 0
      \end{array}
      \endgroup
   \right]
   \;\;\;\;  ,
\end{equation}

\begin{equation}  
   \hat{P}^{(6)}_1
   =
   \!
   \frac{1}{4}
   \!\left[
      \begingroup\SmallColSep
      \begin{array}{r r r r}
                  1 & \bar{1} & \bar{1} &          1 \\
         \bar{1} &          1 &          1 & \bar{1} \\
         \bar{1} &          1 &          1 & \bar{1} \\
                  1 & \bar{1} & \bar{1} &          1
      \end{array}
      \endgroup
   \right]
   \,
   ,
   \,\,
   \hat{P}^{(6)}_2
   =
   \!
   \frac{1}{4}
   \!\left[
      \begingroup\SmallColSep
      \begin{array}{r r r r}
         1 & 1 & 1 & 1 \\
         1 & 1 & 1 & 1 \\
         1 & 1 & 1 & 1 \\
         1 & 1 & 1 & 1
      \end{array}
      \endgroup
   \right]
   \,
   ,
   \,\,
   \hat{P}^{(6)}_3
   =
   \!
   \frac{1}{2}
   \!\left[
      \begingroup\SmallColSep
      \begin{array}{r r r r}
                  1 &          0 &          0 & \bar{1} \\
                  0 &          0 &          0 &          0 \\
                  0 &          0 &          0 &          0 \\
         \bar{1} &          0 &          0 &          1
      \end{array}
      \endgroup
   \right]
   \,
   ,
   \,\,
   \hat{P}^{(6)}_4
   =
   \!
   \frac{1}{2}
   \!\left[
      \begingroup\SmallColSep
      \begin{array}{r r r r}
                  0 &            0 &            0 &          0 \\
                  0 &            1 &   \bar{1} &          0 \\
                  0 &   \bar{1} &            1 &          0 \\
                  0 &            0 &            0 &          0
      \end{array}
      \endgroup
   \right]
   \;\;\;\;  ,
\end{equation}
\smallskip

\noindent where $\bar{1}$ stands for $-1$.\\

\noindent As it can be readily seen, all $\hat{P}^{(1)}_i$ are orthogonal to each other and $\sum_{i=1}^4 \hat{P}^{(1)}_i = \hat{1}$. The same is true for $\hat{P}^{(6)}_i$, which means that $\mathcal{C}^{(1)}$ and $\mathcal{C}^{(6)}$ are the maximal contexts. Their column and null spaces are:\smallskip

\begin{equation}  
   \mathrm{ran}(\hat{P}^{(1)}_1)
   =
   \left\{
      \left[
      \begin{array}{r}
          0  \\
          0  \\
          0  \\
          a
      \end{array}
      \!\!
      \right]
      \!
      :\,
      a \in \mathbb{R}
      \right\}
      \,
      ,
      \,\,
      \mathrm{ran}(\neg\hat{P}^{(1)}_1)
      =
      \left\{
         \left[
         \begin{array}{r}
             b  \\
             c  \\
             d  \\
             0
         \end{array}
         \!\!
         \right]
         \!
         :\,
         b,c,d \in \mathbb{R}
         \right\}
   \;\;\;\;  ,
\end{equation}

\begin{equation}  
   \mathrm{ran}(\hat{P}^{(1)}_2)
   =
   \left\{
      \left[
      \begin{array}{r}
          0  \\
          a  \\
          0  \\
          0
      \end{array}
      \!\!
      \right]
      \!
      :\,
      a \in \mathbb{R}
      \right\}
      \,
      ,
      \,\,
      \mathrm{ran}(\neg\hat{P}^{(1)}_2)
      =
      \left\{
         \left[
         \begin{array}{r}
             b  \\
             0  \\
             c  \\
             d
         \end{array}
         \!\!
         \right]
         \!
         :\,
         b,c,d \in \mathbb{R}
         \right\}
   \;\;\;\;  ,
\end{equation}

\begin{equation}  
   \mathrm{ran}(\hat{P}^{(1)}_3)
   =
   \left\{
      \left[
      \begin{array}{r}
          0  \\
          0  \\
          a  \\
          0
      \end{array}
      \!\!
      \right]
      \!
      :\,
      a \in \mathbb{R}
      \right\}
      \,
      ,
      \,\,
      \mathrm{ran}(\neg\hat{P}^{(1)}_3)
      =
      \left\{
         \left[
         \begin{array}{r}
            -c  \\
             b  \\
             c  \\
             d
         \end{array}
         \!\!
         \right]
         \!
         :\,
         b,c,d \in \mathbb{R}
         \right\}
   \;\;\;\;  ,
\end{equation}

\begin{equation}  
   \mathrm{ran}(\hat{P}^{(1)}_4)
   =
   \left\{
      \left[
      \begin{array}{r}
          a  \\
          0  \\
         -a  \\
          0
      \end{array}
      \!\!
      \right]
      \!
      :\,
      a \in \mathbb{R}
      \right\}
      \,
      ,
      \,\,
      \mathrm{ran}(\neg\hat{P}^{(1)}_4)
      =
      \left\{
         \left[
         \begin{array}{r}
             c  \\
             b  \\
             c  \\
             d
         \end{array}
         \!\!
         \right]
         \!
         :\,
         b,c,d \in \mathbb{R}
         \right\}
   \;\;\;\;  ;
\end{equation}
\smallskip

\begin{equation}  
   \mathrm{ran}(\hat{P}^{(6)}_1)
   =
   \left\{
      \left[
      \begin{array}{r}
          a  \\
         -a  \\
         -a  \\
          a
      \end{array}
      \!\!
      \right]
      \!
      :\,
      a \in \mathbb{R}
      \right\}
      \,
      ,
      \,\,
      \mathrm{ran}(\neg\hat{P}^{(6)}_1)
      =
      \left\{
         \left[
         \begin{array}{r}
             b+c-d  \\
                    b   \\
                    c   \\
                    d
         \end{array}
         \!\!
         \right]
         \!
         :\,
         b,c,d \in \mathbb{R}
         \right\}
   \;\;\;\;  ,
\end{equation}

\begin{equation}  
   \mathrm{ran}(\hat{P}^{(6)}_2)
   =
   \left\{
      \left[
      \begin{array}{r}
          a  \\
          a  \\
          a  \\
          a
      \end{array}
      \!\!
      \right]
      \!
      :\,
      a \in \mathbb{R}
      \right\}
      \,
      ,
      \,\,
      \mathrm{ran}(\neg\hat{P}^{(6)}_2)
      =
      \left\{
         \left[
         \begin{array}{r}
              -b-c-d  \\
                     b  \\
                     c  \\
                     d
         \end{array}
         \!\!
         \right]
         \!
         :\,
         b,c,d \in \mathbb{R}
         \right\}
   \;\;\;\;  ,
\end{equation}

\begin{equation}  
   \mathrm{ran}(\hat{P}^{(6)}_3)
   =
   \left\{
      \left[
      \begin{array}{r}
          a  \\
          0  \\
          0  \\
         -a
      \end{array}
      \!\!
      \right]
      \!
      :\,
      a \in \mathbb{R}
      \right\}
      \,
      ,
      \,\,
      \mathrm{ran}(\neg\hat{P}^{(6)}_3)
      =
      \left\{
         \left[
         \begin{array}{r}
             d  \\
             b  \\
             c  \\
             d
         \end{array}
         \!\!
         \right]
         \!
         :\,
         b,c,d \in \mathbb{R}
         \right\}
   \;\;\;\;  ,
\end{equation}

\begin{equation}  
   \mathrm{ran}(\hat{P}^{(6)}_4)
   =
   \left\{
      \left[
      \begin{array}{r}
          0  \\
          a  \\
         -a  \\
          0
      \end{array}
      \!\!
      \right]
      \!
      :\,
      a \in \mathbb{R}
      \right\}
      \,
      ,
      \,\,
      \mathrm{ran}(\neg\hat{P}^{(6)}_4)
      =
      \left\{
         \left[
         \begin{array}{r}
             b  \\
             c  \\
             c  \\
             d
         \end{array}
         \!\!
         \right]
         \!
         :\,
         b,c,d \in \mathbb{R}
         \right\}
   \;\;\;\;  .
\end{equation}
\smallskip

\noindent Let the system be prepared in the state $|1^{(1)}\rangle$ lying in the column space of the projection operator $\hat{P}^{(1)}_1$ and so in the null spaces of the rest of the projections from the context $\mathcal{C}^{(1)}$, which, in accordance with (\ref{three}) and (\ref{four}), implies\smallskip

\begin{equation}  
   |1^{(1)}\rangle
   \in
   \left\{
      \left[
      \begin{array}{r}
          0  \\
          0  \\
          0  \\
          a
      \end{array}
      \!\!
      \right]
   \right\}
   \,
   \implies
   \,
   v_{|1^{(1)}\rangle}(\hat{P}^{(1)}_1)
   =
   1
   \;\;\;\;  ,
\end{equation}

\begin{equation}  
   |1^{(1)}\rangle
   \in
   \left\{
      \begin{array}{r}
         \left\{
            \left[
            \begin{array}{r}
                b  \\
                c  \\
                0  \\
                d
            \end{array}
            \!\!
            \right]
         \right\}
         \,
         \implies
         \,
         v_{|1^{(1)}\rangle}(\hat{P}^{(1)}_2)
         =
         0
         \\
         \left\{
            \left[
            \begin{array}{r}
               -c  \\
                b  \\
                c  \\
                d
            \end{array}
            \!\!
            \right]
         \right\}
         \,
         \implies
         \,
         v_{|1^{(1)}\rangle}(\hat{P}^{(1)}_3)
         =
         0
         \\
         \left\{
            \left[
            \begin{array}{r}
                c  \\
                b  \\
                c  \\
                d
            \end{array}
            \!\!
            \right]
         \right\}
         \,
         \implies
         \,
         v_{|1^{(1)}\rangle}(\hat{P}^{(1)}_4)
         =
         0
   \end{array}
   \right.
   \;\;\;\;  .
\end{equation}
\smallskip

\noindent In consequence, $\sum_{i=1}^4 v_{|1^{(1)}\rangle}(\hat{P}^{(1)}_i) = 1$.\\

\noindent Consider the intersections\smallskip

\begin{equation}  
   \mathrm{ran}(\hat{P}^{(1)}_1)
   \cap
   \mathrm{ran}(\hat{P}^{(6)}_1)
   =
   \left\{
      \left[\!\!
      \begin{array}{r}
          0  \\
          0  \\
          0  \\
          a
      \end{array}
      \!\!
      \right]
   \right\}
   \cap
   \left\{
      \left[\!\!
      \begin{array}{r}
          a  \\
         -a  \\
         -a  \\
          a
      \end{array}
      \!\!
      \right]
   \right\}
   =
   \{0\}
   \;\;\;\;  ,
\end{equation}

\begin{equation}  
   \mathrm{ran}(\hat{P}^{(1)}_1)
   \cap
   \mathrm{ran}(\hat{P}^{(6)}_2)
   =
   \left\{
      \left[\!\!
      \begin{array}{r}
          0  \\
          0  \\
          0  \\
          a
      \end{array}
      \!\!
      \right]
   \right\}
   \cap
   \left\{
      \left[\!\!
      \begin{array}{r}
          a  \\
          a  \\
          a  \\
          a
      \end{array}
      \!\!
      \right]
   \right\}
   =
   \{0\}
   \;\;\;\;  ,
\end{equation}

\begin{equation}  
   \mathrm{ran}(\hat{P}^{(1)}_1)
   \cap
   \mathrm{ran}(\hat{P}^{(6)}_3)
   =
   \left\{
      \left[\!\!
      \begin{array}{r}
          0  \\
          0  \\
          0  \\
          a
      \end{array}
      \!\!
      \right]
   \right\}
   \cap
   \left\{
      \left[\!\!
      \begin{array}{r}
          a  \\
          0  \\
          0  \\
         -a
      \end{array}
      \!\!
      \right]
   \right\}
   =
   \{0\}
   \;\;\;\;  ,
\end{equation}

\begin{equation}  
   \mathrm{ran}(\hat{P}^{(1)}_1)
   \cap
   \mathrm{ran}(\hat{P}^{(6)}_4)
   =
   \left\{
      \left[\!\!
      \begin{array}{r}
          0  \\
          0  \\
          0  \\
          a
      \end{array}
      \!\!
      \right]
   \right\}
   \cap
   \left\{
      \left[\!\!
      \begin{array}{r}
          0  \\
          a  \\
         -a  \\
          0
      \end{array}
      \!\!
      \right]
   \right\}
   =
   \{0\}
   \;\;\;\;  .
\end{equation}
\smallskip

\noindent Because every one of these intersections is the zero subspace, $\mathrm{ran}(\hat{P}^{(1)}_1)$ is orthogonal to every $\mathrm{ran}(\hat{P}^{(6)}_i)$ and, hence, $v_{|1^{(1)}\rangle}(\neg\hat{P}^{(1)}_1) = v_{|1^{(1)}\rangle}(\hat{P}^{(6)}_i) = 0$. However, this leads to a contradiction, namely, $v_{|1^{(1)}\rangle}(\sum_{i=1}^4 \hat{P}^{(6)}_i) = 1 = \sum_{i=1}^4 v_{|1^{(1)}\rangle}(\hat{P}^{(6)}_i) = 0$.\\

\noindent Now, consider additional intersections:\smallskip

\begin{equation}  
   \mathrm{ran}(\hat{P}^{(1)}_1)
   \cap
   \mathrm{ran}(\neg\hat{P}^{(6)}_1)
   =
   \left\{
      \left[\!\!
      \begin{array}{r}
          0  \\
          0  \\
          0  \\
          a
      \end{array}
      \!\!
      \right]
   \right\}
   \cap
   \left\{
      \left[\!\!
      \begin{array}{r}
          b+c-d  \\
                b  \\
                c  \\
                d
      \end{array}
      \!\!
      \right]
   \right\}
   =
   \{0\}
   \;\;\;\;  ,
\end{equation}

\begin{equation}  
   \mathrm{ran}(\hat{P}^{(1)}_1)
   \cap
   \mathrm{ran}(\neg\hat{P}^{(6)}_2)
   =
   \left\{
      \left[\!\!
      \begin{array}{r}
          0  \\
          0  \\
          0  \\
          a
      \end{array}
      \!\!
      \right]
   \right\}
   \cap
   \left\{
      \left[\!\!
      \begin{array}{r}
         -b-c-d  \\
                b  \\
                c  \\
                d
      \end{array}
      \!\!
      \right]
   \right\}
   =
   \{0\}
   \;\;\;\;  ,
\end{equation}

\begin{equation}  
   \mathrm{ran}(\hat{P}^{(1)}_1)
   \cap
   \mathrm{ran}(\neg\hat{P}^{(6)}_3)
   =
   \left\{
      \left[\!\!
      \begin{array}{r}
          0  \\
          0  \\
          0  \\
          a
      \end{array}
      \!\!
      \right]
   \right\}
   \cap
   \left\{
      \left[\!\!
      \begin{array}{r}
          d  \\
          b  \\
          c  \\
          d
      \end{array}
      \!\!
      \right]
   \right\}
   =
   \{0\}
   \;\;\;\;  ,
\end{equation}

\begin{equation}  
   \mathrm{ran}(\hat{P}^{(1)}_1)
   \cap
   \mathrm{ran}(\neg\hat{P}^{(6)}_4)
   =
   \left\{
      \left[\!\!
      \begin{array}{r}
          0  \\
          0  \\
          0  \\
          a
      \end{array}
      \!\!
      \right]
   \right\}
   \cap
   \left\{
      \left[\!\!
      \begin{array}{r}
          b  \\
          c  \\
          c  \\
          d
      \end{array}
      \!\!
      \right]
   \right\}
   =
   \left\{
      \left[\!\!
      \begin{array}{r}
          0  \\
          0  \\
          0  \\
          a
      \end{array}
      \!\!
      \right]
   \right\}
   \,
   \implies
   \,
   \left\{
      \left[\!\!
      \begin{array}{r}
          0  \\
          0  \\
          0  \\
          a
      \end{array}
      \!\!
      \right]
   \right\}
   \subseteq
   \left\{
      \left[\!\!
      \begin{array}{r}
          b  \\
          c  \\
          c  \\
          d
      \end{array}
      \!\!
      \right]
   \right\}
   \;\;\;\;  .
\end{equation}
\smallskip

\noindent From these intersections it follows that $v_{|1^{(1)}\rangle}(\hat{P}^{(6)}_{i \neq 4}) = 1$ and $v_{|1^{(1)}\rangle}(\hat{P}^{(6)}_{4}) = 0$. But this leads to another contradiction, namely, $v_{|1^{(1)}\rangle}(\sum_{i=1}^4 \hat{P}^{(6)}_i) = 1 = \sum_{i=1}^4 v_{|1^{(1)}\rangle}(\hat{P}^{(6)}_i) = 3$.\\

\noindent So, if the system is prepared in the pure state $|1^{(1)}\rangle$ in which $v_{|1^{(1)}\rangle}(\hat{P}^{(1)}_1) = 1$, both $v_{|1^{(1)}\rangle}(\hat{P}^{(6)}_{i \neq 4}) = 1$ and $v_{|1^{(1)}\rangle}(\hat{P}^{(6)}_{i \neq 4}) = 0$ lead to contradictions. Hence, $\hat{P}^{(6)}_{i \neq 4}$ must be value indefinite under $v_{|1^{(1)}\rangle}$, that is,\smallskip

\begin{equation}  
   |1^{(1)}\rangle
   \notin
   \left\{
      \begin{array}{r}
         \mathrm{ran}(\hat{P}^{(6)}_{i \neq 4})
      \\
         \mathrm{ran}(\neg\hat{P}^{(6)}_{i \neq 4})
      \end{array}
   \right.
   \iff
   v_{|1^{(1)}\rangle}(\hat{P}^{(6)}_{i \neq 4})
   \notin
   \{0,1\}
   \;\;\;\;  .
\end{equation}
\smallskip

\section{Interpretation of the Kochen-Specker theorem}  

\noindent The failure of being a total function for the evaluation relation $v_{|1^{(1)}\rangle}:\, \{\hat{P}^{(6)}_i\} \rightarrow \{0,1\}$ can be described by way of \textit{the truth-value gaps}, namely,\smallskip

\begin{equation} \label{five} 
   |\Omega\rangle
   \notin
   \left\{
      \begin{array}{r}
         \mathrm{ran}(\hat{P}_{\diamond})
      \\
         \mathrm{ran}(\neg\hat{P}_{\diamond})
      \end{array}
   \right.
   \iff
   \left\{
      v_{|\Omega\rangle}
      (
         \hat{P}_\diamond
      )
   \right\}
   =
   \varnothing
   \;\;\;\;  .
\end{equation}
\smallskip

\noindent This expression means that in a state $|\Omega\rangle$ not residing in the column or null space of a projection operator $\hat{P}_{\diamond}$, a proposition $\diamond$ associated with $\hat{P}_{\diamond}$ has no truth-value at all, i.e., $\{ {[\![ \diamond ]\!]}_{|\Omega\rangle} \}= \varnothing$. A semantics defined by set of these truth-value gaps in conjunction with the valuations (\ref{three}) and (\ref{four}) is \textit{gappy and yet two-valued}. Accordingly, it can be called \textit{a supervaluationist semantics} (for details of such semantics see \cite{Varzi, Keefe} and also \cite{Bolotin17, Bolotin18}).\\

\noindent This semantics is, in general, \textit{not truth-functional}: Thus, according to (\ref{three}), (\ref{four}) and (\ref{five}), in any admissible state of the system, the truth-value assignment function assigns the value of the truth to the sum of the projection operators $\hat{P}_\diamond$ in a maximal context $\mathcal{C}$, even though there is a state $|\Omega\rangle$ where at least one of these projection operators has no truth-value:\smallskip

\begin{equation}  
   \left.
      \begin{array}{r}
         v_{|\Omega\rangle}(\hat{1}) = 1
      \\
         \{v_{|\Omega\rangle}(\hat{P}_\diamond)\} = \varnothing
      \end{array}
   \right\}
   \,
   \implies
   \,
   \begin{array}{r}
      v_{|\Omega\rangle}
         \left(
            \sum_{\hat{P}_\diamond \in \mathcal{C}}
               \hat{P}_\diamond
            \right)
            =
            1
   \\
      \left\{
         \sum_{\hat{P}_\diamond \in \mathcal{C}}
            v_{|\Omega\rangle}(\hat{P}_\diamond)
      \right\}
      =
      \varnothing
   \end{array}
   \;\;\;\;  .
\end{equation}
\smallskip

\noindent For example, the values $v_{|1^{(1)}\rangle}(\hat{P}^{(6)}_{i \neq 4})$ are nonexistent within the supervaluationist semantics, and so $\{\sum_{i=1}^4 v_{|1^{(1)}\rangle}(\hat{P}^{(6)}_i) \} = \varnothing$.\\

\noindent Alternatively, the failure of the principle of bivalence can be described using \textit{a multivalued semantics} in which projection operators $\hat{P}$ may have more than two values, specifically,\smallskip

\begin{equation}  
   v_C
   :
   \{
      \hat{P}
   \}
   \rightarrow
   \mathcal{V}_N
   \;\;\;\;  ,
\end{equation}
\smallskip

\noindent where $\mathcal{V}_N$ denotes a set of truth-values whose cardinality is $N > 2$ and whose upper and lower bounds are 1 (that represents ``true'' or ``absolutely true'') and 0 (that represents ``false'' or ``absolutely false''), respectively.\\

\noindent To accomplish that, instead of the truth-value gaps (\ref{five}) one can introduce the following valuation\smallskip

\begin{equation} \label{six} 
   |\Omega\rangle
   \notin
   \left\{
      \begin{array}{r}
         \mathrm{ran}(\hat{P}_{\diamond})
      \\
         \mathrm{ran}(\neg\hat{P}_{\diamond})
      \end{array}
   \right.
   \iff
   v_{|\Omega\rangle}
   \!
   (
      \hat{P}_\diamond
   )
   =
   \langle\Omega|\hat{P}_\diamond|\Omega\rangle
   \in
   \{x \in \mathbb{R}\,|\,0<x<1\}
   \;\;\;\;  ,
\end{equation}
\smallskip

\noindent where the function $v_{|\Omega\rangle}$ is determined by the probability $\mathbb{P}[ {[\![ \diamond ]\!]}_{|\Omega\rangle}\!\! =\! 1 ] = \langle\Omega|\hat{P}_\diamond|\Omega\rangle$. As it is said in \cite{Pykacz95, Pykacz15} the value $v_{|\Omega\rangle}(\hat{P}_\diamond)$ represents the degree to which the projection operator $\hat{P}_\diamond \in \mathcal{C}$ has the value 1 in the state $|\Omega\rangle$. Because $\langle\Omega|\hat{P}_\diamond|\Omega\rangle \in [0,1]$, a semantics defined by the set of the valuations (\ref{three}), (\ref{four}) and (\ref{six}) is i\textit{nfinite-valued}.\\

\noindent For example, in the infinite-valued semantics, the values $v_{|1^{(1)}\rangle}(\hat{P}^{(6)}_{i \neq 4})$ are defined by $v_{|1^{(1)}\rangle}(\hat{P}^{(6)}_1) = v_{|1^{(1)}\rangle}(\hat{P}^{(6)}_2) = \textonequarter$ and $v_{|1^{(1)}\rangle}(\hat{P}^{(6)}_3) = \textonehalf$ inferring $\sum_{i=1}^4 v_{|1^{(1)}\rangle}(\hat{P}^{(6)}_i) = 1$. This shows that despite its value indefiniteness under the bivaluation, the context $\mathcal{C}^{(6)}$ is value definite under the infinite-valued assignment (\ref{six}).\\

\noindent Hence, only within the supervaluationist semantics, the negation of the sentence (a) (i.e., the principle of bivalence) can be interpreted as a sign that the verification of the ``gappy'' (i.e., having no truth-values) propositions  must result in the \textit{ex nihilo} creation of the bivalent values of these propositions.\\

\noindent While on the contrary, in the many-valued semantics, the failure of bivalence implies that non-classical (i.e., different from 1 and 0) truth-values must exist before the verification and they become bivalent as a result of the verification (thus, a bivalent semantic merely emerges at the end of the verification process). In such a sense, one may say that the measurements (verifications) produce the output that yield pre-existing elements of physical reality.\\

\noindent For that reason, the Kochen-Specker theorem (along with its localized variant) cannot justify the belief that quantum mechanics is indeterministic, that is, that there are no hidden variables (or not-yet-verified truth values of the propositions) determining somehow the outcome of a measurement (verification) in advance. This theorem only shows that if those hidden variables were to exist, they would have to comply with a logic which does not obey the principle of bivalence.\\

\section*{Acknowledgment}

\noindent The author would like to express his appreciation to the anonymous referee whose attentive comments helped to improve this paper considerably.\\

\bibliographystyle{References}
\bibliography{QC_ref}

\begin{thebibliography}{10}
\expandafter\ifx\csname urlstyle\endcsname\relax
  \providecommand{\doi}[1]{doi:\discretionary{}{}{}#1}\else
  \providecommand{\doi}{doi:\discretionary{}{}{}\begingroup
  \urlstyle{rm}\Url}\fi

\bibitem{Abbott12}
A.~Abbott, C.~Calude, J.~Conder, and K.~Svozil.
\newblock Strong {K}ochen-{S}pecker theorem and incomputability of quantum
  randomness.
\newblock \emph{Phys. Rev. A}, 86(062109):1--11, 2012.

\bibitem{Kochen}
S.~Kochen and E.~Specker.
\newblock The problem of hidden variables in quantum mechanics.
\newblock \emph{J. Math. Mech.}, 17(1):59--87, 1967.

\bibitem{Peres}
A.~Peres.
\newblock Two simple proofs of the {K}ochen-{S}pecker theorem.
\newblock \emph{Phys. A: Math. Gen.}, 24:L175--L178, 1991.

\bibitem{Abbott15}
A.~Abbott, C.~Calude, and K.~Svozil.
\newblock A variant of the {K}ochen-{S}pecker theorem localizing value
  indefiniteness.
\newblock \emph{J. Math. Phys.}, 56(102201):1--17, 2015.

\bibitem{Beziau}
J.-Y. B{$\grave{\mathrm{e}}$}ziau.
\newblock Bivalence, {E}xcluded {M}iddle and {N}on {C}ontradiction.
\newblock In L.~Behounek, editor, \emph{The {L}ogica {Y}earbook 2003}, pages
  73--84. Academy of Sciences, Prague, 2003.

\bibitem{Cabello}
A.~Cabello, J.~Estebaranz, and G.~Garcia-{A}lcaine.
\newblock Bell-{K}ochen-{S}pecker theorem: {A} proof with 18 vectors.
\newblock \emph{Phys. Letters A}, 212:183--187, 1996.

\bibitem{Varzi}
A.~Varzi.
\newblock Supervaluationism and {I}ts {L}ogics.
\newblock \emph{Mind}, 116:633--676, 2007.

\bibitem{Keefe}
R.~Keefe.
\newblock \emph{Theories of {V}agueness}.
\newblock Cambridge University Press, Cambridge, 2000.

\bibitem{Bolotin17}
A.~Bolotin.
\newblock Quantum supervaluationism.
\newblock \emph{J. Math. Phys.}, 58(12):122106--1--7, 2017.
\newblock \doi{10.1063/1.5008374}.

\bibitem{Bolotin18}
A.~Bolotin.
\newblock Truth {V}alues of {Q}uantum {P}henomena.
\newblock \emph{Int. J. Theor. Phys.}, 57(7):2124--2132, 2018.
\newblock \doi{10.1007/s10773-018-3737-z}.

\bibitem{Pykacz95}
J.~Pykacz.
\newblock Quantum {L}ogic as {P}artial {I}nfinite-{V}alued {{\L}}ukasiewicz
  {L}ogic.
\newblock \emph{Int. J. Theor. Phys.}, 34(8):1697--1710, 1995.

\bibitem{Pykacz15}
J.~Pykacz.
\newblock \emph{Quantum {P}hysics, {F}uzzy {S}ets and {L}ogic. {S}teps
  {T}owards a {M}any-{V}alued {I}nterpretation of {Q}uantum {M}echanics}.
\newblock Springer, 2015.

\end{thebibliography}

\end{document}